\documentclass[11pt]{article}
\usepackage{amsmath, amsfonts, amssymb}
\usepackage{amsthm} 
\usepackage{hyperref}

\newcommand{\set}[1]{{\lbrace{#1}\rbrace}}

\newcommand{\bra}[1]{{\left\langle #1 \right|}}
\newcommand{\ket}[1]{{\left| #1 \right\rangle}}
\newcommand{\tensor}{\otimes}
\newcommand{\Tensor}{\bigotimes}

\newcommand{\CC}{\mathbb{C}}
\newcommand{\MM}{\mathbb{M}}
\newcommand{\eps}{\varepsilon}

\newcommand{\calA}{\mathcal{A}}
\newcommand{\calE}{\mathcal{E}}
\newcommand{\calM}{\mathcal{M}}

\DeclareMathOperator{\EE}{\mathbb{E}}
\DeclareMathOperator{\Tr}{tr}

\theoremstyle{plain}   
\newtheorem{definition}{Definition}[section]
\newtheorem{theorem}[definition]{Theorem}

\newtheorem{corollary}[definition]{Corollary}
\theoremstyle{definition}

\begin{document}



\title{Single-shot security for one-time memories in the isolated qubits model}

\author{Yi-Kai Liu\\
Applied and Computational Mathematics Division\\
National Institute of Standards and Technology (NIST)\\
Gaithersburg, MD, USA\\
yi-kai.liu@nist.gov}
\date{\today}
\maketitle

\abstract{
One-time memories (OTM's) are simple, tamper-resistant cryptographic devices, which can be used to implement sophisticated functionalities such as one-time programs.  Can one construct OTM's whose security follows from some physical principle?  This is not possible in a fully-classical world, or in a fully-quantum world, but there is evidence that OTM's can be built using ``isolated qubits'' --- qubits that cannot be entangled, but can be accessed using adaptive sequences of single-qubit measurements.

Here we present new constructions for OTM's using isolated qubits, which improve on previous work in several respects:  they achieve a stronger ``single-shot'' security guarantee, which is stated in terms of the (smoothed) min-entropy; they are proven secure against adversaries who can perform arbitrary local operations and classical communication (LOCC); and they are efficiently implementable.  

These results use Wiesner's idea of conjugate coding, combined with error-correcting codes that approach the capacity of the $q$-ary symmetric channel, and a high-order entropic uncertainty relation, which was originally developed for cryptography in the bounded quantum storage model.


}


\section{Introduction}

\textit{One-time memories} (OTM's) are a simple type of tamper-resistant cryptographic hardware.  An OTM has the following behavior:  a user Alice can write two messages $s$ and $t$ into the OTM, and then give the OTM to another user Bob; Bob can then choose to read either $s$ or $t$ from the OTM, but he can only learn one of the two messages, not both.  A single OTM is not especially exciting by itself, but when many OTM's are combined in an appropriate way, they can be used to implement \textit{one-time programs}, which are a powerful form of secure computation \cite{GKR,goyal,bellare,broadbent}.  (Roughly speaking, a one-time program is a program that can be run exactly once, on an input chosen by the user.  After running once, the program ``self-destructs,'' and it never reveals any information other than the output of the computation.)

Can one construct OTM's whose security follows from some physical principle?  At first glance, the answer seems to be ``no.''  OTM's cannot exist in a fully classical world, because information can always be copied without destroying it.  One might hope to build OTM's in a quantum world, where the no-cloning principle limits an adversary's ability to copy an unknown quantum state.  However, this is also impossible, because an OTM can be used to perform oblivious transfer with information-theoretic security, which is ruled out by various ``no-go'' theorems \cite{nogo1,nogo2,nogo3,nogo4}.  

One way around these no-go theorems is to try to construct protocols that are secure against restricted classes of quantum adversaries, e.g., adversaries who can only perform $k$-local measurements \cite{salvail}, or adversaries who only have bounded or noisy quantum storage \cite{bounded05,bounded06,bounded07,noisy07,tight-eur,all-but-one}.  More recently, Liu has proposed a construction for OTM's in the \textit{isolated qubits model} \cite{liu}, where the adversary is only allowed to perform local operations and classical communication (LOCC).  That is, the adversary can perform single-qubit quantum operations, including single-qubit measurements, and can make adaptive choices based on the classical information returned by these measurements; but the adversary cannot perform entangling operations on sets of two or more qubits.  (Honest parties are also restricted to LOCC operations.)  The isolated qubits model is motivated by recent experimental work using solid-state qubits, such as nitrogen vacancy (NV) centers; see \cite{liu} for a more complete discussion of this model, and \cite{pastawski} for earlier work on implementing quantum money using NV centers.
\footnote{
Note that the devices constructed in \cite{liu}, and in this paper, are more precisely described as \textit{leaky} OTM's, because they can leak additional information to the adversary.  It is not known whether such leaky OTM's are sufficient to construct one-time programs as defined in \cite{GKR}.  We will discuss this issue in Section \ref{sec-outlook}; for now, we will simply refer to our devices as OTM's.
}

In this paper we show a new construction and security analysis for OTM's in the isolated qubits model, which improves on the results of \cite{liu} in several respects.  First, we show a stronger ``single-shot'' security guarantee, which is stated in terms of the (smoothed) min-entropy \cite{renner,renner-wolf}.  This shows that a constant fraction of the message bits remain hidden from the adversary.  This stronger statement is necessary for most cryptographic applications; note that the previous results of \cite{liu} were not sufficient, as they used the Shannon entropy.  

Second, we prove security against general LOCC adversaries, who can perform arbitrary measurements (including weak measurements), and can measure each qubit multiple times.  This improves on the results of \cite{liu}, which only showed security against 1-pass LOCC adversaries that use 2-outcome measurements.  Our new security proof is based solely on the definition of the isolated qubits model, without any additional assumptions.  

Third, we show a construction of OTM's that is efficiently implementable, i.e., programming and reading out the OTM can be done in polynomial time.  This improves on the construction in \cite{liu}, which was primarily an information-theoretic result, using random error-correcting codes that did not allow efficient decoding.  (In fact, our new construction is quite flexible, and does not depend heavily on the choice of a particular error-correcting code.  Our OTM's can be constructed using any code that satisfies two simple requirements:  the code must be linear over $GF(2)$, and it must approach the capacity of the $q$-ary symmetric channel.  We show one such code in this paper; several more sophisticated constructions are known \cite{bky,shok,bms}.)  

We will describe our OTM construction in the following section.  Here, we briefly comment on some related work.  Note that OTM's cannot make use of standard techniques such as privacy amplification.  This is because OTM's are non-interactive and asynchronous:  all of the communication between Alice and Bob occurs at the beginning, while the adversary can wait until later to attack the OTM.  (To do privacy amplification, Alice would have to first force the adversary to take some action, and then send one more message to Bob.  This trick is very natural in protocols for quantum key distribution and oblivious transfer, but it is clearly impossible in the case of an OTM.)  As we will see below, the security of our OTM's follows from rather different arguments.  (A similar issue was studied recently in \cite{all-but-one}, albeit with a weaker, non-adaptive adversary.)

In addition, it is a long-standing open problem to prove strong upper-bounds on the power of LOCC operations.  Previous results in this area include demonstrations of ``nonlocality without entanglement'' \cite{nlwe99} (see \cite{nlwe12} for a recent survey), and constructions of data-hiding states \cite{qdh02,eggeling02,hqd02,mpdhqi05}.  Our OTM's are not directly comparable to these earlier results, as the security requirements for our OTM's are quite different.  

\subsection{Our construction}
\label{sec-proof-techniques}

We now describe our OTM construction, which is based on Wiesner's idea of conjugate coding \cite{wiesner}.  Our OTM will store two messages $s,t \in \set{0,1}^\ell$, and will use $n\lg q$ qubits, where $q$ is a (large) power of 2.  Let $C:\: \set{0,1}^\ell \rightarrow \set{0,1}^{n\lg q}$ be any error-correcting code that satisfies the following two requirements:  $C$ is linear over $GF(2)$, and $C$ approaches the capacity of the $q$-ary symmetric channel $\calE_q$ with error probability $p_e := \tfrac{1}{2} - \tfrac{1}{2q}$ (where the channel treats each block of $\lg q$ bits as a single $q$-ary symbol).  Note that, when $q$ is large, the capacity of the channel $\calE_q$ is roughly $1-p_e$, which is roughly $\tfrac{1}{2}$, so we have $n\lg q \approx 2\ell$.

Given two messages $s$ and $t$, let $C(s)$ and $C(t)$ be the corresponding codewords, and view each codeword as $n$ blocks consisting of $\lg q$ bits.  We prepare the qubits in the OTM as follows.  For each $i = 1,2,\ldots,n$, 
\begin{itemize}
\item Let $\gamma_i \in \set{0,1}$ be the outcome of a fair and independent coin toss.
\item If $\gamma_i=0$, prepare the $i$'th block of qubits in the standard basis state corresponding to the $i$'th block of $C(s)$.
\item If $\gamma_i=1$, prepare the $i$'th block of qubits in the Hadamard basis state corresponding to the $i$'th block of $C(t)$.
\end{itemize}

To recover the first message $s$, we measure every qubit in the standard basis, which yields a string of measurement outcomes $z \in \set{0,1}^{n\lg q}$, and then we run the decoding algorithm for $C$.  To recover the second message $t$, we measure every qubit in the Hadamard basis, then follow the same procedure.  It is easy to see that all of these procedures require only single-qubit state preparations and single-qubit measurements, which are allowed in the isolated qubits model.
\footnote{
We note in passing that Winter's ``gentle measurement lemma'' \cite{gentle} does not imply an attack on this OTM using LOCC operations.
The idea behind the gentle measurement lemma is that, if there is a \textit{nondestructive} measurement that recovers $s$ with high probability, and there is a similar measurement for $t$, then one can perform both measurements, and recover both $s$ and $t$ with high probability.  

However, the LOCC measurement that recovers $s$ is \textit{destructive}, as is the LOCC measurement for $t$.  This is because one has to perform a projective measurement on each qubit, obtain a string of classical measurement outcomes, and then run the classical decoding algorithm for $C$.  In order to use the gentle measurement lemma, one would have to perform these measurements nondestructively, which would require running the decoding algorithm for $C$ on a superposition of many different inputs; and this would require entangling operations.
}

(We remark that this OTM construction uses blocks of qubits, rather than individual qubits as in \cite{wiesner} and \cite{liu}.  That is, we set $q$ large, instead of using $q=2$.  This difference seems to help our security proof, although it is not clear whether it affects the actual security of the scheme.)

We now sketch the proofs of correctness and security for this OTM.  With regard to correctness, note that an honest player who wanted to learn $s$ will obtain measurement outcomes that have the same distribution as the output of the $q$-ary symmetric channel $\calE_q$ acting on $C(s)$; hence the decoding algorithm will return $s$.  A similar argument holds for $t$.

To prove security, we consider adversaries that make \textit{separable} measurements (which include LOCC measurements as a special case).  The basic idea is to consider the distribution of the messages $s$ and $t$, conditioned on one particular measurement outcome $z$ obtained by the adversary.  Since the adversary is separable, the corresponding POVM element $M_z$ will be a tensor product of single-qubit operators $\Tensor_{a=1}^{n\lg q} R_a$ (up to normalization).  Now, one can imagine a fictional adversary that measures the qubits one at a time, and happens to observe this same string of single-qubit measurement outcomes $R_1,R_2,\ldots,R_{n\lg q}$.  This event leads to the same conditional distribution of $s$ and $t$.  But the fictional adversary is easier to analyze, because it is non-adaptive, it measures each qubit only once, and the measurements can be done in arbitrary order.

Now, our proof will be based on the following intuition.  In order to learn both messages $s$ and $t$, the adversary will want to determine the basis choices $\gamma = (\gamma_1,\gamma_2,\ldots,\gamma_n)$, so that he will know which blocks of qubits should be measured in the standard basis, and which blocks of qubits should be measured in the Hadamard basis.  The choice of the code $C$ is crucial to prevent the adversary from doing this; for instance, if the adversary could predict some of the bits in the codewords $C(s)$ and $C(t)$, he could then measure the corresponding qubits, and gain some information about which bases were used to prepare them.  (Note moreover that the adversary has full knowledge of $C$, before he measures any of the qubits.)  We will argue that certain properties of the code $C$ prevent the adversary from learning these basis choices $\gamma$ perfectly, and that this in turn limits the adversary's knowledge of the messages $s$ and $t$.  

Since $C$ is a linear code over $GF(2)$, it has a generator matrix $G$, which has rank $\ell$.  Thus there must exist a subset of $\ell$ bits of the codeword $C(s)$ that look uniformly random, assuming the message $s$ was chosen uniformly at random; and a similar statement holds for $C(t)$.  Now, let $A$ be the subset of $\ell$ qubits that encode these bits of $C(s)$ and $C(t)$.  We can imagine that the fictional adversary happens to measure these qubits \textit{first}.  Therefore, during these first $\ell$ steps, the fictional adversary learns nothing about which bases had been used to prepare the state, i.e., the basis choices $\gamma$ are independent of the fictional adversary's measurement outcomes.  

One can then show that the conditional distribution of $s$ and $t$ after these first $\ell$ steps of the fictional adversary is related to the distribution of measurement outcomes when the state $\Tensor_{a\in A} R_a$ is measured in a random basis.  This kind of situation has been studied previously, in connection with cryptography in the bounded quantum storage model.  In particular, we can use a high-order entropic uncertainty relation from \cite{tight-eur} to show a lower-bound on the smoothed min-entropy of this distribution.  We then use trivial bounds to analyze the remaining $n\lg q - \ell$ steps of the fictional adversary.  Roughly speaking, we get a bound of the form:
\begin{equation}
\label{eqn-snail}
H_\infty^\eps(S,T|Z) \gtrsim \tfrac{1}{2}\,\ell,
\end{equation}
for any separable adversary (where $Z$ denotes the adversary's measurement outcome).  Thus, while the OTM may leak some information, it still hides a constant fraction of the bits of the messages $s$ and $t$.  For more details, see Section \ref{sec-porpoise}.

Finally, we show one construction of a code $C$ that satisfies the above requirements and is efficiently decodable.  The basic idea is to fix some $q_0 < q$, first encode the messages $s$ and $t$ using a random linear code $C_0:\: \set{0,1}^\ell \rightarrow \set{0,1}^{n\lg q_0}$, then encode each block of $\lg q_0$ bits using a fixed linear code $C_1:\: \set{0,1}^{\lg q_0} \rightarrow \set{0,1}^{\lg q}$.  The code $C_1$ is used to detect the errors made by the $q$-ary symmetric channel; these corrupted blocks of bits are then treated as erasures, and we can decode $C_0$ by solving a linear system of equations, which can be done efficiently.  Moreover, choosing $C_0$ to be a random linear encode ensures that, with high probability, $C$ approaches the capacity of the $q$-ary symmetric channel.  For more details, see Section \ref{sec-seal}.

\subsection{Outlook}
\label{sec-outlook}

The results of this paper can be summarized as follows:  we construct OTM's based on conjugate coding, which achieve a fairly strong (``single-shot'') notion of security, are secure against general LOCC adversaries, and can be implemented efficiently.  These results are a substantial improvement on previous work \cite{liu}.  

We view these results as a first step in a broader research program that aims to develop practical implementations of isolated qubits, one-time memories, and ultimately one-time programs.  We now comment briefly on some different aspects of this program.

Experimental realization of isolated qubits is quite challenging, though there has been recent progress in this direction \cite{silicon,diamond}.  Broadly speaking, isolated qubits seem to be at an intermediate level of difficulty, somewhere between photonic quantum key distribution (which already exists as a commercial product), and large-scale quantum computers (which are still many years in the future).

Working with quantum devices in the lab also raises the question of fault-tolerance:  can our OTM's be made robust against minor imperfections in the qubits?  We believe this can be done, by slightly modifying our OTM construction:  we would use a slightly noisier channel to describe the imperfect measurements made by an honest user, and we would choose the error-correcting code $C$ accordingly.  The proof of security would still hold against LOCC adversaries who can make perfect measurements.  There is plenty of ``slack'' in the security bounds, to allow this modification to the OTM's.

In addition, one may wonder whether our OTM's are secure against so-called ``$k$-local'' adversaries \cite{salvail}, which can perform entangled measurements on small numbers of qubits (thus going outside the isolated qubits model).  There is some reason to be optimistic about this:  while we have mainly discussed separable adversaries in this paper, our security proof actually works for a larger set of adversaries, who can generate entanglement among some of the qubits, but are still separable across the partition defined by the subset $A$ (as described in the proof).  Also, from a physical point of view, $k$-local adversaries are quite natural.  In particular, even when one can perform entangling operations on pairs of qubits, it may be hard to entangle large numbers of qubits, due to error accumulation.  

Finally, let us turn to the construction of one-time programs.  Because our OTM's leak some information, it is not clear whether they are sufficient to construct one-time programs.  There are a couple of approaches to this problem.  On one hand, one can try to strengthen the security proof, perhaps by proving constraints on the \textit{types} of information that an LOCC adversary can extract from the OTM.  We conjecture that, when our OTM's are used to build one-time programs as in \cite{GKR}, the specific information that is relevant to the security of the one-time program does in fact remain hidden from an LOCC adversary.  

On the other hand, one can try to strengthen the OTM constructions, in order to eliminate the leakage.  As noted previously, standard privacy amplification (e.g., postprocessing using a randomness extractor) does not work in this setting, because the adversary also knows the seed for the extractor.  However, there are other ways of solving this problem, for instance by assuming the availability of a random oracle, or by using something similar to leakage-resilient encryption \cite{AGV,naor-segev} (but with a different notion of leakage, where the ``leakage function'' is restricted to use only LOCC operations, but is allowed access to side-information).  


\section{Preliminaries}

\subsection{Notation}

For any natural number $n$, let $[n]$ denote the set $\set{1,2,\ldots,n}$.  
Let $\lg(x) = \log_2(x)$ denote the logarithm with base 2.

For any random variable $X$, let $P_X$ be the probability density function of $X$, that is, $P_X(x) = \Pr[X=x]$.  Likewise, define $P_{X|Y}(x|y) = \Pr[X=x|Y=y]$, etc.  For any event $\calE$, define $P_{\calE X}$ to be the probability density function of $X$ smoothed by $\calE$, that is $P_{\calE X}(x) = \Pr[X=x \text{ and } \calE \text{ occurs}]$.  

We say that $C$ is a binary code with codeword length $n$ and message length $k$ if $C$ is a subset of $\set{0,1}^n$ with cardinality $2^k$.  We say that $C$ has minimum distance $d = \min_{x,y\in C} d_H(x,y)$, where $d_H(\cdot,\cdot)$ denotes the Hamming distance.  

We say that $C$ is a binary linear code if $C$ is a linear subspace of $GF(2)^n$.  (Note, $GF(2)$ and $\set{0,1}$ denote the same set, but we will write $GF(2)$ in situations where we use arithmetic operations.)  In this case, there exists a matrix $G \in GF(2)^{k\times n}$, such that the map $x \mapsto x^T G$ is a bijection from $GF(2)^k$ to the code subspace $C$.  We will overload the notation and use $C$ to denote the map $x \mapsto x^T G$; then the codewords consist of the strings $C(x)$ for all $x \in GF(2)^k$.

\subsection{The $q$-ary symmetric channel}

The $q$-ary symmetric channel with error probability $p_e$ acts as follows:  given an input $x \in GF(q)$, it returns an output $y \in GF(q)$, with conditional probabilities $\Pr(y|x) = 1-p_e$ (if $y=x$) and $\Pr(y|x) = p_e/(q-1)$ (if $y \neq x$).
The capacity of this channel, measured in $q$-ary symbols per channel use, is given by \cite{shok}:
\begin{equation} \label{eqn-llama}
\begin{split}
L(p_e) &= 1 + (1-p_e)\log_q(1-p_e) + p_e\log_q(p_e) - p_e\log_q(q-1) \\
 &= 1 - \frac{h_2(p_e)}{\lg q} - p_e\frac{\lg(q-1)}{\lg q} 
  \geq 1 - \frac{1}{\lg q} - p_e,
\end{split}
\end{equation}
where $h_2(\cdot)$ is the binary entropy function.  

\subsection{LOCC adversaries and separable measurements}

An LOCC adversary is an adversary that uses only local operations and classical communication (LOCC).  Here, ``local operations'' consist of quantum operations on single qubits, and ``classical communication'' refers to the adversary's ability to choose each single-qubit operation adaptively, depending on classical information, such as measurement outcomes, that were obtained from previous single-qubit operations.  However, the adversary is not allowed to make adaptive choices that depend on quantum information, or perform entangling operations on multiple qubits.

Formally, an LOCC adversary can be described as follows.  Consider a system of $n$ qubits.  The adversary makes a sequence of steps, labelled by $i=1,2,3,\ldots$.  At step $i$, the adversary chooses one of the qubits $q_i \in [n]$, and performs a general quantum measurement $\calM_i$ on that qubit; this returns a measurement outcome, which is described by a classical random variable $Z_i$.  The adversary's choices of $q_i$ and $\calM_i$ can depend on $Z_1,Z_2,\ldots,Z_{i-1}$.  Also, note that the adversary can perform weak measurements, and can measure the same qubit multiple times.  Finally the adversary discards the qubits, and outputs the sequence of measurement outcomes $Z_1,Z_2,Z_3,\ldots$.  

A POVM measurement $\calM = \set{M_z \;|\; z=1,2,3,\ldots}$ is called \textit{separable} if every POVM element $M_z$ can be written as a tensor product of single-qubit operators.  It is easy to see that any LOCC adversary can be simulated by a separable measurement, i.e., for any LOCC adversary $\calA$, there exists a separable POVM measurement $\calM$, such that for every quantum state $\rho$, the output of $\calM$ acting on $\rho$ has the same distribution as the output of $\calA$ acting on $\rho$ \cite{horodecki}.

\subsection{Leaky OTM's}

We will use the following definition of a leaky OTM \cite{liu}.
\begin{definition}
Fix some class of adversary strategies $\MM$, some leakage parameter $\delta \in [0,1]$, and some failure probability $\eps \in [0,1]$.  A \emph{leaky one-time memory (leaky OTM)} with parameters $(\MM,\delta,\eps)$ is a device that has the following behavior.  Suppose that the device is programmed with two messages $s$ and $t$ chosen uniformly at random in $\set{0,1}^\ell$; and let $S$ and $T$ be the random variables containing these messages.  Then:
\begin{enumerate}
\item Correctness:  There exists an honest strategy $\calM^{(1)} \in \MM$ that interacts with the device and recovers the message $s$ with probability $\geq 1-\eps$.  Likewise, there exists an honest strategy $\calM^{(2)} \in \MM$ that recovers the message $t$ with probability $\geq 1-\eps$.  
\item Leaky security:  For every strategy $\calM \in \MM$, if $Z$ is the random variable containing the classical information output by $\calM$, then $H_\infty^\eps(S,T|Z) \geq (1-\delta)\ell$.
\end{enumerate}
\end{definition}

Here $H_\infty^\eps$ is the smoothed conditional min-entropy, which is defined as follows \cite{renner,renner-wolf}:  
\begin{equation}
H_\infty^\eps(X|Y) = \max_{\calE:\; \Pr(\calE) \geq 1-\eps} \min_{x,y} \Bigl[ -\lg \bigl[ P_{\calE X|Y}(x|y) \bigr] \Bigr], 
\end{equation}
where the maximization is over all events $\calE$ (defined by the conditional probabilities $P_{\calE|XY}$) such that $\Pr(\calE) \geq 1-\eps$.  Observe that a lower-bound of the form $H_\infty^\eps(X|Y) \geq h$ implies that there exists an event $\calE$ with $\Pr(\calE) \geq 1-\eps$ such that, for all $x$ and $y$, $\Pr[\calE,X=x|Y=y] \leq 2^{-h}$.  

The definition of a leaky OTM is weaker than that of an ideal OTM in two important respects:  it assumes that the messages $s$ and $t$ are chosen uniformly at random, independent of all other variables; and it allows the adversary to obtain partial information about both $s$ and $t$, so long as the adversary still has $(1-\delta)k$ bits of uncertainty (as measured by the smoothed min-entropy).  We suspect that this \textit{definition} of a leaky OTM is not strong enough to construct one-time programs (although we conjecture that our actual \textit{constructions} of OTM's in Sections \ref{sec-porpoise} and \ref{sec-seal} are, in fact, strong enough for this purpose).

\subsection{Uncertainty relations for the min-entropy}

We will use an uncertainty relation from \cite{tight-eur}, with a slight modification to describe quantum systems that consist of many non-identical subsystems:
\begin{theorem}\label{thm-eur}
Consider a quantum system with Hilbert space $\Tensor_{i=1}^{\ell_0} \CC^{d_i}$, i.e., the system can be viewed as a collection of $\ell_0$ subsystems, where the $i$'th subsystem has Hilbert space dimension $d_i$.  

For each $i \in [\ell_0]$, let $B_i$ be a finite collection of orthonormal bases for $\CC^{d_i}$, and suppose that these bases satisfy the following uncertainty relation:  for every quantum state $\rho$ on $\CC^{d_i}$, $|B_i|^{-1} \sum_{\omega \in B_i} H(P_\omega) \geq h_i$, where $P_\omega$ is the distribution of measurement outcomes when $\rho$ is measured in basis $\omega$.  

Now let $\rho$ be any quantum state over $\Tensor_{i=1}^{\ell_0} \CC^{d_i}$, let $\Theta = (\Theta_1,\ldots,\Theta_{\ell_0})$ be chosen uniformly at random from $B_1 \times \cdots \times B_{\ell_0}$, and let $X = (X_1,\ldots,X_{\ell_0})$ be the measurement outcome when $\rho$ is measured in basis $\Theta$ (i.e., each $X_i$ is the outcome of measuring subsystem $i$ in basis $\Theta_i$).  

Then, for any $\tau > 0$, and any $\lambda_1, \ldots, \lambda_{\ell_0} \in (0,\tfrac{1}{2})$, we have:
\begin{equation}
H_\infty^\eps(X|\Theta) \geq -\tau + \sum_{i=1}^{\ell_0} (h_i - \lambda_i),
\end{equation}
where $\eps \leq \exp(-2\tau^2/c)$, and $c = \sum_{i=1}^{\ell_0} 16\bigl(\lg\frac{|B_i|d_i}{\lambda_i}\bigr)^2$.
\end{theorem}
The proof is essentially the same as in \cite{tight-eur}; it uses a martingale argument and Azuma's inequality, but it allows the martingale to have different increments at each step.  

In addition, we will use the following chain rule for the smoothed min-entropy \cite{renner-wolf}:
\begin{equation}\label{eqn-seagull}
H_\infty^{\eps+\eps'}(X|Y) > H_\infty^\eps(X,Y) - H_0(Y) - \lg(\tfrac{1}{\eps'}).
\end{equation}


\section{One-time memories}
\label{sec-porpoise}

We now show the correctness and security of the OTM construction described in Section \ref{sec-proof-techniques}.  Recall that this OTM uses $n\lg q$ qubits, stores two messages of length $\ell$, and uses an error-correcting code $C$.  We will show how to set $n$ and $q$, and how to choose the code $C$.  

Let us introduce some notation.  We view the code $C$ as a function $C: \set{0,1}^{\ell} \rightarrow \set{0,1}^{n\lg q}$.  We view each codeword $x \in \set{0,1}^{n\lg q}$ as a sequence of $n$ blocks, where each block is a binary string of length $\lg q$.  We write the codeword as $x = (x_{ij})_{i\in[n], j\in[\lg q]}$, and we write the $i$'th block as $x_i = (x_{ij})_{j\in[\lg q]}$.  Finally, let $H$ be the Hadamard gate acting on a single qubit.  

We now prepare the qubits in the OTM as follows.  For each $i = 1,2,\ldots,n$, 
\begin{itemize}
\item Let $\gamma_i \in \set{0,1}$ be the outcome of a fair and independent coin toss.
\item If $\gamma_i=0$, prepare the $i$'th block of qubits in the state $\ket{C(s)_i}$.
\item If $\gamma_i=1$, prepare the $i$'th block of qubits in the state $H^{\tensor (\lg q)} \ket{C(t)_i}$.
\end{itemize}
To recover the first message $s$, we measure every qubit in the standard basis, which yields a string of measurement outcomes $z \in \set{0,1}^{n\lg q}$, and then we run the decoding algorithm for $C$.  To recover the second message $t$, we measure every qubit in the Hadamard basis, obtain a string of measurement outcomes $z$, and again run the decoding algorithm for $C$.  

We will prove the following general theorem, which works for any code $C$ that satisfies certain properties:
\begin{theorem}\label{thm-turkey-hawk}
Let $q\geq 2$ be any power of 2.  Let $\calE_q$ be the $q$-ary symmetric channel with error probability $p_e = (1/2) - (1/2q)$.  Let $\ell \geq 1$ and $n\geq 1$, and let $C: \set{0,1}^\ell \rightarrow \set{0,1}^{n\lg q}$ be any error-correcting code that satisfies the following two requirements:
\begin{enumerate}
\item $C$ can transmit information reliably over the channel $\calE_q$ (where the channel treats each block of $\lg q$ bits as a single $q$-ary symbol).  
\item $C$ is a linear code over $GF(2)$.
\end{enumerate}
Then the above OTM stores two messages $s,t \in \set{0,1}^\ell$, and has the following properties:
\begin{enumerate}
\item The OTM behaves correctly for honest parties.  
\item For any small constants $0 < \lambda \ll \tfrac{1}{2}$, $0 < \tau_0 \ll 1$, and $0 < \delta \ll 1$, the following statement holds.  Suppose the messages $s$ and $t$ are chosen independently and uniformly at random in $\set{0,1}^\ell$.  For any separable adversary,\footnote{Note that this includes LOCC adversaries as a special case.} we have the following security bound:
\begin{equation}
\begin{split}
H&_\infty^{\delta+\eps}(S,T|Z) \\
&\geq \Bigl( (\tfrac{1}{2}-\lambda) 
- 4\tau_0\, ( 1 + \tfrac{1}{\sqrt{\lg q}} (1+\lg\tfrac{1}{\lambda}) ) 
+ (2-\tfrac{1}{\alpha}) \Bigr) \cdot \ell - \lg\tfrac{1}{\delta} \\
&\gtrsim \Bigl( \tfrac{1}{2} + (2-\tfrac{1}{\alpha}) \Bigr) \cdot \ell.
\end{split}
\end{equation}
Here $S$ and $T$ are the random variables describing the two messages, $Z$ is the random variable representing the adversary's measurement outcome, we have $\eps \leq \exp(-2\tau_0^2 \ell/\lg q)$, and $\alpha = \ell/(n\lg q)$ is the rate of the code $C$.
\end{enumerate}
\end{theorem}

Note that, to get a strong security bound, one must use a code $C$ whose rate $\alpha$ is large.  It is useful to ask, then, how large $\alpha$ can be.  Let $L_q$ denote the capacity of the channel $\calE_q$, measured in $q$-ary symbols per channel use.  Using a good code $C$, we can hope to have rate $\alpha \approx L_q$.  Moreover, $L_q$ is lower-bounded by:
\begin{equation}
L_q \geq 1 - \tfrac{1}{\lg q} - p_e = \tfrac{1}{2} - \tfrac{1}{\lg q} + \tfrac{1}{2q} \approx \tfrac{1}{2},
\end{equation}
which is nearly tight when $q$ is large.  So we can hope to have $\alpha \approx \tfrac{1}{2}$, in which case our security bound becomes:
\begin{equation}
H_\infty^{\delta+\eps}(S,T|Z)
\gtrsim \tfrac{1}{2}\,\ell.
\end{equation}


\subsection{Correctness for honest parties}

We first show the ``correctness'' part of Theorem \ref{thm-turkey-hawk}.  Without loss of generality, suppose we want to recover the first message $s$.  (A similar argument applies if we want to recover the second message $t$.)  Let $z \in \set{0,1}^{n\lg q}$ be the string of measurement outcomes obtained by measuring each qubit in the standard basis.  Observe that $z$ is the output of a $q$-ary symmetric channel $\calE_q$ with error probability $p_e = (1/2) - (1/2q)$, acting on the string $C(s) \in \set{0,1}^{n\lg q}$ (viewed as a sequence of $n$ symbols in $GF(q)$).  Since the code $C$ can transmit information reliably over this channel, it follows that we can recover $s$.  


\subsection{Security against separable adversaries}

\begin{figure}
\begin{center}
\input{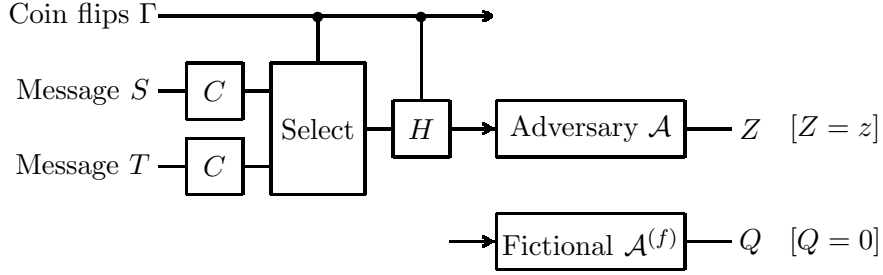}
\end{center}
\caption{OTM with separable adversary $\calA$, and ``fictional'' adversary $\calA^{(f)}$.  In the proof, we will analyze the distributions of $S$ and $T$ conditioned on the events $Z=z$ and $Q=0$.}
\label{fig-adversary}
\end{figure}

We now show the ``security'' part of Theorem \ref{thm-turkey-hawk}.  Let us first introduce some notation (see Figure \ref{fig-adversary}).  Suppose the OTM is programmed with two messages $s$ and $t$ that are chosen independently and uniformly at random in $\set{0,1}^\ell$.  Let $S$ and $T$ be the random variables representing these messages.  Let $\Gamma$ be the random variable representing the coin flips $\gamma = (\gamma_1,\ldots,\gamma_n)$ used in programming the OTM.  $C$ denotes the error-correcting code, which maps $\set{0,1}^\ell$ to $\set{0,1}^{n\lg q}$.  ``Select'' is an operation that maps $\set{0,1}^{n\lg q} \times \set{0,1}^{n\lg q}$ to $\set{0,1}^{n\lg q}$, depending on the value of $\Gamma$, as follows:
\begin{equation}
\text{Select}(x,y)_{i,j} = 
\begin{cases}
x_{i,j} &\text{if } \Gamma_i=0,\\
y_{i,j} &\text{if } \Gamma_i=1,
\end{cases}
\qquad \text{for all $i\in[n]$, $j\in[\lg q]$}.
\end{equation}
``Select'' outputs a string of $n\lg q$ classical bits, which are converted into $n\lg q$ qubits (in the standard basis states $\ket{0}$ and $\ket{1}$).  $H$ denotes a Hadamard gate controlled by the value of $\Gamma$; that is, for each $i\in[n]$ and $j\in[\lg q]$, if $\Gamma_i=1$, then $H$ is applied to the $(i,j)$'th qubit.

Fix any separable adversary $\calA$, let $L$ be the number of possible outcomes that can be observed by the adversary, and let $\calM = \set{M_z \;|\; z\in[L]}$ be the separable POVM measurement performed by the adversary.  Let $Z$ be the random variable representing the adversary's output; so $Z$ takes values in $[L]$.

Fix some small constant $\delta > 0$.  We say that a measurement outcome $z\in[L]$ is ``negligible'' if $\Pr[Z=z] \leq (\delta/2^{n\lg q}) \Tr(M_z)$.  Note that the probability of observing any of these ``negligible'' measurement outcomes is small:  
\begin{equation}\label{eqn-greenmold}
\Pr[Z \text{ is ``negligible''}] = \sum_{z \text{ ``negl.''}} \Pr[Z=z] 
\leq (\delta/2^{n\lg q}) \sum_{z \text{ ``negl.''}} \Tr(M_z) \leq \delta.
\end{equation}

The proof will proceed as follows:  for all messages $s,t \in \set{0,1}^\ell$, and for all measurement outcomes $z\in[L]$ that are not ``negligible,'' we will upper-bound $\Pr[S=s,T=t|Z=z]$.  This will imply a lower-bound on $H_\infty^\delta(S,T|Z)$, which is what we desire.  

\subsubsection{A fictional adversary}
\label{sec-dragonfly}

We begin by fixing some measurement outcome $z\in[L]$ that is not ``negligible.''  Since the adversary performed a separable measurement, we can write the corresponding POVM element $M_z$ as a tensor product of single-qubit operators.  In particular, we can write $M_z = \Tr(M_z) \Tensor_{i=1}^n \Tensor_{j=1}^{\lg q} R_{ij}$, where each $R_{ij}$ is a single-qubit operator, positive semidefinite, with trace 1.

We now construct a fictional adversary $\calA^{(f)}$, which we will use in the proof.  The fictional adversary acts in the following way:  for each qubit $(i,j) \in [n] \times [\lg q]$, it performs the POVM measurement $\set{R_{ij}, I-R_{ij}}$ on qubit $(i,j)$, which yields a binary measurement outcome $Q_{ij}$ (where $Q_{ij}=0$ corresponds to the POVM element $R_{ij}$, and $Q_{ij}=1$ corresponds to $I-R_{ij}$).  Let us write the vector of measurement outcomes as $Q = (Q_{ij})_{i\in[n],\, j\in[\lg q]}$, which takes values in $\set{0,1}^{n\lg q}$.  Let 0 denote the vector $(0,0,\ldots,0) \in \set{0,1}^{n\lg q}$.

Intuitively, the event $Q=0$ (in an experiment using the fictional adversary) corresponds to the event $Z=z$ (in an experiment using the real adversary).  More precisely, for any $s,t \in \set{0,1}^\ell$, we have 
\begin{equation}\label{eqn-bluemold}
\begin{split}
P_{ST|Z}&(s,t|z) 
 = \frac{P_{Z|ST}(z|s,t) P_{ST}(s,t)}{P_Z(z)} \\
&= \frac{P_{Q|ST}(0|s,t) \Tr(M_z) P_{ST}(s,t)}{P_Q(0) \Tr(M_z)} 
 = P_{ST|Q}(s,t|0).
\end{split}
\end{equation}
We will proceed by upper-bounding $P_{ST|Q}(s,t|0)$ (with the fictional adversary); this will imply an upper-bound on $P_{ST|Z}(s,t|z)$ (with the real adversary).

\subsubsection{Properties of the codewords $C(S)$ and $C(T)$}
\label{sec-hummingbird}

Recall that the messages $S$ and $T$ are independently and uniformly distributed in $GF(2)^\ell$.  Now consider the codewords $C(S)$ and $C(T)$.  We claim that there exists a subset of $\ell$ coordinates of $C(S)$ and $C(T)$ that are independently and uniformly distributed in $GF(2)^\ell$.  

To see this, recall that $C$ is a linear code over $GF(2)$.  Hence the encoding operation $C:\: GF(2)^\ell \rightarrow GF(2)^{n\lg q}$ can be written in the form $C(x) = x^T G$ for some matrix $G \in GF(2)^{\ell \times n\lg q}$.  Since the codewords $C(x)$ are all distinct, the matrix $G$ must have row-rank $\ell$.  Hence the column-rank of $G$ must also be $\ell$, so there exists a subset of $\ell$ columns of $G$ that are linearly independent over $GF(2)$.  Let us denote this subset by $A \subset [n]\times[\lg q]$, $|A| = \ell$.

Now look at those coordinates of $C(S)$ and $C(T)$ that correspond to the subset $A$; we write these as $C(S)_A = (C(S)_{ij})_{(i,j)\in A}$ and $(C(T)_{ij})_{(i,j)\in A}$.  It follows that $C(S)_A$ and $C(T)_A$ are independently and uniformly distributed in $GF(2)^\ell$.  

\subsubsection{Behavior of the fictional adversary on the subset of qubits $A$}

We now analyze the behavior of the fictional adversary on those qubits belonging to the subset $A$.  Without loss of generality, we can assume that the fictional adversary measures the qubits in the subset $A$ first, and then measures the remaining qubits in the subset $([n]\times[\lg q]) \setminus A$.  (This follows because the fictional adversary is \textit{non-adaptive}, in that it makes all its decisions about what measurements to perform, before seeing any of the results of the measurements; and because all of the measurements commute with one another, since each measurement only involves a single qubit.)  

For convenience, let $B = ([n]\times[\lg q]) \setminus A$.  Let $Q_A = (Q_{ij})_{(i,j)\in A}$ denote the measurement outcomes of the qubits in the subset $A$, and let $Q_B = (Q_{ij})_{(i,j)\in B}$ denote the measurement outcomes of the qubits in the subset $B$.  

We claim that the OTM's coin tosses $\Gamma$, conditioned on the event $Q_A=0$, are still uniformly distributed in $\set{0,1}^n$.  This is a fairly straightforward calculation; 
see Appendix \ref{app-poplar} for details.

\subsubsection{Using the uncertainty relation}

We will upper-bound these probabilities $P_{ST|\Gamma Q_A}(s,t|\gamma,0)$, using an entropic uncertainty relation.  The basic idea is to consider another experiment, where one runs the OTM and the fictional adversary ``backwards'' in time.  This experiment can be analyzed using the uncertainty relation in Theorem \ref{thm-eur} (originally due to \cite{tight-eur}).  

\begin{figure}
\input{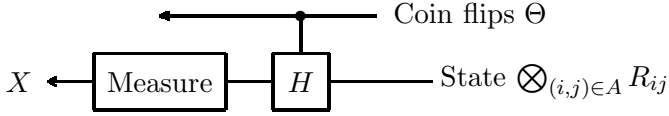}
\caption{In order to understand the behavior of the fictional adversary, conditioned on the event $Q_A=0$, we consider an analogous experiment, where the state $\Tensor_{(i,j) \in A} R_{ij}$ is measured in a random basis.  We will analyze this using an entropic uncertainty relation.}
\label{fig-backwards}
\end{figure}

We now describe this new experiment (see Figure \ref{fig-backwards}).  One prepares the quantum state $\Tensor_{(i,j) \in A} R_{ij}$, one chooses a uniformly random sequence of measurement bases $\Theta = (\Theta_1,\ldots,\Theta_n)$ (where $\Theta_i=0$ denotes the standard basis and $\Theta_i=1$ denotes the Hadamard basis), and then one measures each qubit $(i,j) \in A$ in the basis $\Theta_i$ to get a measurement outcome $X_{ij}$ (which can be either 0 or 1).  

Intuitively, the state $\Tensor_{(i,j) \in A} R_{ij}$ corresponds to the fictional adversary's measurement outcome $Q_A=0$, the random bases $\Theta$ correspond to the OTM's coin flips $\Gamma$, and the measurement outcomes $X$ correspond to those bits $C(S)_A$ and $C(T)_A$ used in the OTM.  (Note that the OTM's coin flips $\Gamma$ are uniformly distributed, even when one conditions on the event $Q_A=0$, as shown in the previous section.)

To make this intuition precise, we will first show that:
\begin{equation}\label{eqn-juniper2}
H_\infty^\eps(S,T|\Gamma, Q_A=0) = H_\infty^\eps(X|\Theta) + \ell.
\end{equation}
(See Appendix \ref{app-aspen} for details.)  
Then note that conditioning on $\Gamma$ can only reduce the entropy, hence we have:
\footnote{
Note that, for all $s$ and $t$, $P_{\calE' ST | Q_A}(s,t|0) = \sum_\gamma P_{\calE' ST | \Gamma Q_A}(s,t|\gamma, 0) P_{\Gamma|Q_A}(\gamma|0) \leq 2^{-\ell} 2^{-h}$.  This implies (\ref{eqn-juniper3a}).
}
\begin{equation}\label{eqn-juniper3a}
H_\infty^\eps(S,T|Q_A=0) \geq H_\infty^\eps(X|\Theta) + \ell.
\end{equation}
We then use Theorem \ref{thm-eur} to show a lower-bound on $H_\infty^\eps(X|\Theta)$; 
see Appendix \ref{app-spruce} for details.  

\subsubsection{Combining all the pieces}

The fictional adversary's complete sequence of measurement outcomes is denoted by $Q = (Q_A,Q_B)$.  So far we have analyzed the adversary's actions on those qubits belonging to the subset $A$, and we have shown a lower-bound on $H_\infty^\eps(S,T|Q_A=0)$.  Now, we will show a lower-bound on $H_\infty^\eps(S,T|Q=0)$.  To do this, we bound the adversary's actions on the subset $B$ in a more-or-less trivial way, using the fact that $\Pr[Q=0] = \Pr[Z=z] / \Tr(M_z) \geq \delta / 2^{n\lg q}$, since $z$ was assumed to be ``non-negligible.''  

We will then consider the real adversary, and show a lower-bound on $H_\infty^{\delta+\eps}(S,T|Z)$.  Here we use the following identity that relates the real adversary and the fictional adversary (see equation (\ref{eqn-bluemold})):
\begin{equation}
H_\infty^\eps(S,T|Z=z) = H_\infty^\eps(S,T|Q=0).
\end{equation}
Finally we combine these results to prove the theorem; 
see Appendix \ref{app-cherry} for details.


\section{Efficient implementations of one-time memories}
\label{sec-seal}

In the previous section, we showed that one-time memories can be constructed from any code that approaches the capacity of the $q$-ary symmetric channel, and is linear over $GF(2)$.  In this section, we will construct codes that have these properties, and moreover can be encoded and decoded efficiently.  Using these codes, we will get efficient implementations of one-time memories.

\begin{figure}
\begin{center}
\input{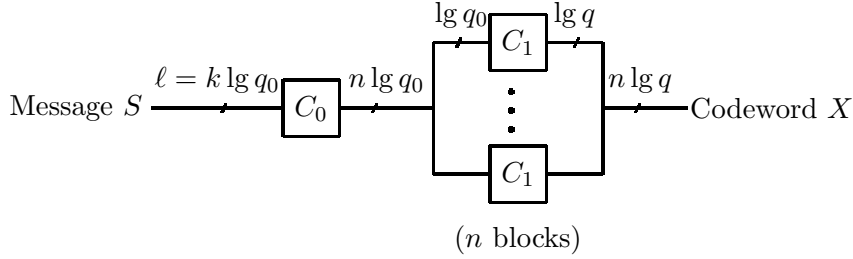}
\end{center}
\caption{Efficient codes for the $q$-ary symmetric channel, based on erasure coding and error detection.}
\label{fig-code}
\end{figure}

There are several known constructions for codes that approach the capacity of the $q$-ary symmetric channel, and are efficiently decodable \cite{bky,shok,bms}.  To illustrate how these techniques can be applied in our setting, we will describe one simple approach, which is based on erasure coding and error detection \cite{shok}.  (See Figure \ref{fig-code}.)  

The basic idea is to take the message $s$, encode it using a code $C_0$ that outputs a string of $q_0$-ary symbols (where $q_0<q$), and then encode each $q_0$-ary symbol using a code $C_1$ that outputs a $q$-ary symbol.  The code $C_1$ is used to detect errors made by the $q$-ary symmetric channel; once detected, these errors can be treated as erasures.  The code $C_0$ is then used to correct these erasures, which is relatively straightforward.  For instance, we can choose $C_0$ to be a random linear code; then we can decode in the presence of erasure errors by solving a linear system of equations, which we can do efficiently.

We now describe the construction in detail.  Let $k\geq 2$ be an integer, let $p_e \in (0,1)$, and choose any small constants $0<\eps \ll 1$, $0<\delta \ll 1$ and $0<\theta \ll 1$.  Define: 
\begin{equation}\label{eqn-def-n}
n = \biggl\lfloor \frac{k}{1-p_e-\theta} \biggr\rfloor,
\end{equation}
\begin{equation}\label{eqn-def-q}
q = 2^c, \quad 
c = \lg q = \bigl\lfloor \tfrac{2}{\delta} \bigl\rfloor \, \bigl\lceil \eps n + \lg(np_e) \bigr\rceil,
\end{equation}
\begin{equation}\label{eqn-def-q0}
q_0 = 2^{c_0}, \quad 
c_0 = \lg q_0 = \bigl\lceil \tfrac{2}{\delta}-2 \bigl\rceil \, \bigl\lceil \eps n + \lg(np_e) \bigr\rceil.
\end{equation}
Note that our setting is slightly unusual, in that we will be constructing codes for the $q$-ary symmetric channel where $q$ is not fixed.  In particular, $\lg q$ (the number of bits used to describe each $q$-ary symbol) grows polynomially with the codeword length $n$, which is proportional to the message length $k$.  

We will construct a code $C:\: \set{0,1}^{k\lg q_0} \rightarrow \set{0,1}^{n\lg q}$ as follows:  
\begin{enumerate}
\item Choose a uniformly random matrix $G_0 \in GF(2)^{k\lg q_0 \times n\lg q_0}$, and define a code $C_0:\: \set{0,1}^{k\lg q_0} \rightarrow \set{0,1}^{n\lg q_0}$ by setting $C_0(s) = s^T G_0$.  
\item Fix any full-rank matrix $G_1 \in GF(2)^{\lg q_0 \times \lg q}$, and define a code $C_1:\: \set{0,1}^{\lg q_0} \rightarrow \set{0,1}^{\lg q}$ by setting $C_1(v) = v^T G_1$.  
\item Define $C(s) = C_1 \circ C_0(s)$, where we view $C_0(s) \in \set{0,1}^{n\lg q_0}$ as a sequence of $n$ blocks of $\lg q_0$ bits, and $C_1$ acts separately on each of these blocks.  Equivalently, we can write $C(s) = s^T G_0 (\bigoplus_{i=1}^n G_1)$, where $\bigoplus_{i=1}^n G_1$ denotes a direct sum of $n$ copies of the matrix $G_1$.
\end{enumerate}

We use the following decoding algorithm:
\begin{enumerate}
\item Given a string $z \in \set{0,1}^{n\lg q}$, write it as a sequence of $n$ blocks of $\lg q$ bits:  $z = (z_{ij})_{i\in[n], j\in[\lg q]}$.
\item For each $i\in[n]$, try to decode the $q$-ary symbol $z_i \in \set{0,1}^{\lg q}$, i.e., try to find some $v \in \set{0,1}^{\lg q_0}$ such that $C_1(v) = z_i$.  Let $b_i$ be the result (or set $b_i = *$ if $z_i$ lies outside the image of $C_1$).  Thus we get a string $b = (b_1,b_2,\ldots,b_n) \in \bigl( \set{0,1}^{\lg q_0} \cup \set{*} \bigr)^n$.
\item Try to decode the string $b$, treating the $*$ symbols as erasures, i.e., try to find some $a \in \set{0,1}^{k\lg q_0}$ such that, for all $i\in[n]$ such that $b_i \neq *$, and for all $j \in [\lg q]$, $C_0(a)_{ij} = b_{ij}$.  If a solution exists, output it; if there are multiple solutions, choose any one of them and output it; otherwise, abort.
\end{enumerate}

Finally, we introduce some more notation.  Let us choose a message (represented by a random variable $S$) uniformly at random in $\set{0,1}^{k\lg q_0}$.  Let $\calE_q$ be the $q$-ary symmetric channel with error probability $p_e$.  We take the message $S$, encode it using the code $C$, transmit it through the channel $\calE_q$, then run the decoding algorithm, and get an estimate of the original message; call this $\hat{S}$.  

We prove the following statement 
(see Appendix \ref{app-saguaro} for details):
\begin{theorem}\label{thm-redtail}
Let $k\geq 2$ be an integer, let $p_e \in (0,1)$, and choose any small constants $0<\eps \ll 1$, $0<\delta \ll 1$ and $0<\theta \ll 1$.  
Let us construct the code $C:\: \set{0,1}^{k\lg q_0} \rightarrow \set{0,1}^{n\lg q}$ as described above.  Then $C$ has the following properties:
\begin{enumerate}
\item With high probability (over the choice of the random matrix $G_0$), $C$ can transmit information reliably over the $q$-ary symmetric channel $\calE_q$ with error probability $p_e$.  

More precisely, choose any small constant $\tau$ such that $0<\tau<\theta$, and choose any large constant $\lambda \gg 1$.  Then, with probability $\geq 1-\frac{1}{\lambda}$ (over the choice of $G_0$), the code $C$ can transmit information over the channel $\calE_q$, and the probability of decoding failure is bounded by:
\begin{equation}
\Pr[\hat{S} \neq S] \leq \lambda \bigl( e^{-2\tau^2 n} + 2^{-\eps n} + 2^{(-n\theta+n\tau+1)\lg q_0} \bigr)
\leq e^{-\Omega(n)}.
\end{equation}

\item $C$ is a linear code over $GF(2)$.

\item $C$ has rate $\alpha := \frac{k\lg q_0}{n\lg q} \geq (1-p_e-\theta) (1-\delta)$.  (Note that this approaches the capacity of the channel $\calE_q$, as shown in equation (\ref{eqn-llama}), when $q$ is large.)

\item The encoding and decoding algorithms for $C$ run in time polynomial in $n\lg q$.  (Also note that $\lg q$ grows at most linearly with $n$, and $n$ is proportional to $k$.)
\end{enumerate}
\end{theorem}

Finally, we can take the code $C$ constructed above (for $p_e=\tfrac{1}{2}$), and combine it with the OTM construction of Theorem \ref{thm-turkey-hawk}, to get the following result:
\begin{corollary}\label{cor-night-hawk}
For any $k \geq 2$, and for any small constant $0<\mu \ll 1$, there exists an OTM construction that stores two messages $s,t \in \set{0,1}^\ell$, where $\ell = \Theta(k^2)$, and has the following properties:
\begin{enumerate}
\item The OTM behaves correctly for honest parties.
\item The OTM can be implemented in time polynomial in $k$.  
\item Let $0 < \delta \ll 1$ be any small constant.  Suppose the messages $s$ and $t$ are chosen independently and uniformly at random in $\set{0,1}^\ell$.  For any separable adversary,\footnote{Note that this includes LOCC adversaries as a special case.} we have the following security bound:
\begin{equation}
\begin{split}
H_\infty^{\delta+\eps}(S,T|Z)
&\geq (\tfrac{1}{2}-\mu) \, \ell - \lg\tfrac{1}{\delta}.
\end{split}
\end{equation}
Here $S$ and $T$ are the random variables describing the two messages, $Z$ is the random variable representing the adversary's measurement outcome, and we have $\eps \leq \exp(-\Omega(k))$.
\end{enumerate}
\end{corollary}


\vskip 11pt
\vskip 11pt

\textbf{Acknowledgements:}
It is a pleasure to thank Serge Fehr, Stephen Jordan, Maris Ozols, Rene Peralta, Eren Sasoglu, Christian Schaffner, Barbara Terhal, Alexander Vardy, and several anonymous reviewers, for helpful suggestions about this work.  Some of these discussions took place at the Schloss Dagstuhl -- Leibniz Center for Informatics.  This paper is a contribution of NIST, an agency of the US government, and is not subject to US copyright.


\appendix

\section{One-time memories}

\subsection{Security against separable adversaries}

\subsubsection{Behavior of the fictional adversary on the subset of qubits $A$}
\label{app-poplar}

We claim that the OTM's coin tosses $\Gamma$, conditioned on the event $Q_A=0$, are still uniformly distributed in $\set{0,1}^n$.  To see this, we first write $P_{Q_A|\Gamma ST}$ as follows:
\begin{equation}\label{eqn-yak}
P_{Q_A|\Gamma ST}(0|\gamma,s,t)
 = \prod_{(i,j) \in A} \Tr(R_{ij}\, \rho(C(s)_{ij}, C(t)_{ij}, \gamma_i)), 
\end{equation}
where for all $x,y,g \in \set{0,1}$, we define the single-qubit state $\rho(x,y,g)$ by 
\begin{equation}
\rho(x,y,g) = 
\begin{cases}
\ket{x}\bra{x} &\text{if } g=0, \\
H\ket{y}\bra{y}H &\text{if } g=1.
\end{cases}
\end{equation}

Next, we write $P_{Q_A|\Gamma}$ as follows:  
\begin{equation}\label{eqn-zebra}
\begin{split}
P_{Q_A|\Gamma}(0|\gamma)
 &= \sum_{s,t \in \set{0,1}^\ell} P_{Q_A|ST\Gamma}(0|s,t,\gamma) P_{ST|\Gamma}(s,t|\gamma) \\
 &= \sum_{s,t \in \set{0,1}^\ell} 4^{-\ell} \prod_{(i,j) \in A} \Tr(R_{ij}\, \rho(C(s)_{ij}, C(t)_{ij}, \gamma_i)) \\
 &= \sum_{a,b \in \set{0,1}^A} 4^{-\ell} \prod_{(i,j) \in A} \Tr(R_{ij}\, \rho(a_{ij}, b_{ij}, \gamma_i)) \\
 &= 4^{-\ell} \prod_{(i,j) \in A} 2\Tr(R_{ij})
 = 2^{-\ell},
\end{split}
\end{equation}
where we used the following facts:  $\Gamma$ is independent of $S$ and $T$; $P_{Q_A|ST\Gamma}(0|s,t,\gamma)$ only depends on those coordinates of $C(s)$ and $C(t)$ corresponding to the subset $A$; these subsets of bits are uniformly distributed in $\set{0,1}^\ell$; and $\Tr(R_{ij}) = 1$.  

Then we can write $P_{Q_A}$ and $P_{\Gamma|Q_A}$ as follows:
\begin{equation}\label{eqn-treefungus}
P_{Q_A}(0) = \sum_{\gamma \in \set{0,1}^n} P_{Q_A|\Gamma}(0|\gamma)\, 2^{-n}
 = 2^{-\ell},
\end{equation}
\begin{equation}\label{eqn-poppy}
P_{\Gamma|Q_A}(\gamma|0) = \frac{P_{Q_A|\Gamma}(0|\gamma)\, 2^{-n}}{P_{Q_A}(0)} = 2^{-n},
\end{equation}
which proves our claim.

We will now calculate the probability distribution of the messages $S$ and $T$, conditioned on the OTM's coin tosses $\Gamma=\gamma$ and the adversary's measurement outcomes $Q_A=0$:
\begin{equation}\label{eqn-minotaur}
\begin{split}
P_{ST|\Gamma Q_A}(s,t|\gamma,0)
 &= \frac{P_{Q_A|ST\Gamma}(0|s,t,\gamma) P_{ST|\Gamma}(s,t|\gamma)}{P_{Q_A|\Gamma}(0|\gamma)} \\
 &= \frac{4^{-\ell} \prod_{(i,j) \in A} \Tr(R_{ij}\, \rho(C(s)_{ij}, C(t)_{ij}, \gamma_i))}{2^{-\ell}} \\
 &= 2^{-\ell} \prod_{(i,j) \in A} \Tr( R_{ij}\, \rho(C(s)_{ij}, C(t)_{ij}, \gamma_i)), 
\end{split}
\end{equation}
where we used (\ref{eqn-yak}), (\ref{eqn-zebra}), and the fact that the $\Gamma$ is chosen independently of $S$ and $T$.

\subsubsection{Using the uncertainty relation}
\label{app-aspen}

We will now show how $H_\infty^\eps(S,T|\Gamma, Q_A=0)$ and $H_\infty^\eps(X|\Theta)$ are related.  We will proceed in several steps.  First, define the following function $\Phi: \set{0,1}^\ell \times \set{0,1}^\ell \times \set{0,1}^n \rightarrow \set{0,1}^A$, 
\begin{equation}
\Phi_{ij}(s,t,\gamma) = 
\begin{cases}
C(s)_{ij} &\text{if } \gamma_i=0, \\
C(t)_{ij} &\text{if } \gamma_i=1
\end{cases}
\quad\quad (\text{for all } (i,j) \in A).
\end{equation}
Define a new random variable $F = \Phi(S,T,\Gamma)$, which takes values in $\set{0,1}^A$.  (Intuitively, $F$ is the output of the ``Select'' function, restricted to those coordinates in the subset $A$.)  We can write the probability distribution of $F$ as follows:
\begin{equation}\label{eqn-gnat}
P_{F|\Gamma Q_A}(f|\gamma,0) = \sum_{(s,t) \;:\; \Phi(s,t,\gamma) = f} P_{ST|\Gamma Q_A}(s,t|\gamma,0).
\end{equation}

How many terms are there in the sum in equation (\ref{eqn-gnat})?  For any fixed $f$ and $\gamma$, define the set $E_{f\gamma} = \set{(s,t) \in \set{0,1}^{2\ell} \;|\; \Phi(s,t,\gamma) = f}$.  Note that we can view $\Phi(s,t,\gamma) = f$ as a set of $\ell$ linear constraints on $s$ and $t$.  In particular, these constraints fix the values of a subset $\set{(i,j) \in A \;|\; \gamma_i = 0}$ of the coordinates of $C(s) = s^T G$, and they fix the values of a subset $\set{(i,j) \in A \;|\; \gamma_i = 1}$ of the coordinates of $C(t) = t^T G$.  Recall from section \ref{sec-hummingbird} that the subset $A$ of the columns of the matrix $G$ is linearly independent.  Hence this set of linear constraints has rank $\ell$, and so the set $E_{f\gamma}$ has size 
\begin{equation}
|E_{f\gamma}| = 2^\ell.
\end{equation}

Also, note that we can write the distribution of $S$ and $T$ (from equation (\ref{eqn-minotaur})) in the following way:
\begin{equation}\label{eqn-spore}
P_{ST|\Gamma Q_A}(s,t|\gamma,0)
 = 2^{-\ell} \prod_{(i,j) \in A} \Tr( R_{ij}\, H^{\gamma_i} \ket{f_{ij}} \bra{f_{ij}} H^{\gamma_i} ), 
 \quad \text{where } f = \Phi(s,t,\gamma).
\end{equation}
Notice that, if we pick $(s,t)$ and $(\tilde{s}, \tilde{t})$ such that $\Phi(s,t,\gamma) = \Phi(\tilde{s}, \tilde{t}, \gamma)$, then $P_{ST|\Gamma Q_A} (s,t|\gamma,0) = P_{ST|\Gamma Q_A} (\tilde{s}, \tilde{t} | \gamma,0)$.  Hence all the terms in the sum in equation (\ref{eqn-gnat}) are identical.  So we can simplify it as follows:
\begin{equation}\label{eqn-morel}
P_{F|\Gamma Q_A}(f|\gamma,0) = \prod_{(i,j) \in A} \Tr( R_{ij}\, H^{\gamma_i} \ket{f_{ij}} \bra{f_{ij}} H^{\gamma_i} ).
\end{equation}
Furthermore, by comparing equations (\ref{eqn-spore}) and (\ref{eqn-morel}), we see that:
\begin{equation}\label{eqn-button}
P_{ST|\Gamma Q_A}(s,t|\gamma,0)
 = 2^{-\ell} P_{F|\Gamma Q_A}(f|\gamma,0), 
 \quad \text{where } f = \Phi(s,t,\gamma).
\end{equation}

Using equations (\ref{eqn-poppy}) and (\ref{eqn-morel}), we can now see that $(F,\Gamma)$ (conditioned on $Q_A=0$) has the same distribution as $(X,\Theta)$.  This implies that: 
\begin{equation}\label{eqn-juniper}
H_\infty^\eps(F|\Gamma, Q_A=0) = H_\infty^\eps(X|\Theta).
\end{equation}
Furthermore, using equation (\ref{eqn-button}), we see that: 
\footnote{
This involves a tedious calculation.  Let $h = H_\infty^\eps(X|\Theta)$.  Equation (\ref{eqn-juniper}) implies that there exists an event $\calE$, with probability $\Pr[\calE | Q_A=0] \geq 1-\eps$, such that for all $f$ and $\gamma$, $P_{\calE F | \Gamma Q_A}(f|\gamma, 0) \leq 2^{-h}$.  This event $\calE$ is defined by the conditional probabilities $\Pr[\calE | F=f, \Gamma=\gamma, Q_A=0]$.  We now define a new event $\calE'$ which has conditional probabilities 
\begin{equation}
\Pr[\calE' | S=s, T=t, \Gamma=\gamma, Q_A=0] = \Pr[\calE | F=\Phi(s,t,\gamma), \Gamma=\gamma, Q_A=0].  
\end{equation}
A straightforward calculation then shows that $\Pr[\calE' | Q_A=0] \geq 1-\eps$, and for all $s$, $t$ and $\gamma$, $P_{\calE' ST | \Gamma Q_A}(s,t|\gamma, 0) \leq 2^{-\ell} 2^{-h}$.  This implies equation (\ref{eqn-juniper2}).
}
\begin{equation}\label{eqn-juniper2}
H_\infty^\eps(S,T|\Gamma, Q_A=0) = H_\infty^\eps(X|\Theta) + \ell.
\end{equation}
Finally, note that conditioning on $\Gamma$ can only reduce the entropy, hence we have:
\footnote{
Note that, for all $s$ and $t$, $P_{\calE' ST | Q_A}(s,t|0) = \sum_\gamma P_{\calE' ST | \Gamma Q_A}(s,t|\gamma, 0) P_{\Gamma|Q_A}(\gamma|0) \leq 2^{-\ell} 2^{-h}$.  This implies (\ref{eqn-juniper3}).
}
\begin{equation}\label{eqn-juniper3}
H_\infty^\eps(S,T|Q_A=0) \geq H_\infty^\eps(X|\Theta) + \ell.
\end{equation}

\subsubsection{Using the uncertainty relation, part 2}
\label{app-spruce}

We now use Theorem \ref{thm-eur} to show a lower-bound on $H_\infty^\eps(X|\Theta)$.  

Recall that the qubits in the OTM are arranged in $n$ blocks, each of size $\lg q$.  The set $A$ describes a subset of these qubits, which are contained in a subset of the blocks.  Let $\Lambda$ be the set of blocks that contain one or more qubits that lie in the set $A$, that is, let $\Lambda = \set{i \in [n] \;|\; \exists j \in [\lg q] \text{ s.t. } (i,j) \in A}$.  For each $i \in \Lambda$, let $A_i$ be the set of qubits in the $i$'th block that lie in the set $A$, that is, let $A_i = \set{j \in [\lg q] \text{ s.t. } (i,j) \in A}$.  So we have $A = \bigcup_{i \in \Lambda} \set{i} \times A_i$.  Let $\ell_0 = |\Lambda|$, and let $\ell_i = |A_i|$; then we have $\ell = \sum_{i \in \Lambda} \ell_i$.  

Using the terminology of Theorem \ref{thm-eur}, we have a quantum system that consists of $\ell_0$ subsystems, where the $i$'th subsystem consists of $\ell_i$ qubits and has dimension $d_i := 2^{\ell_i}$.  For each subsystem $i \in \Lambda$, we have a set $B_i$ that contains two orthonormal bases for $(\CC^2)^{\tensor \ell_i}$, namely the standard basis and the Hadamard basis.  These satisfy the following uncertainty relation \cite{maassen-uffink,wehner-winter}:  for every quantum state $\rho$ on $(\CC^2)^{\tensor \ell_i}$, $|B_i|^{-1} \sum_{\omega \in B_i} H(P_\omega) \geq \ell_i/2 =: h_i$, where $P_\omega$ is the distribution of measurement outcomes when $\rho$ is measured in basis $\omega$.  

Now let $\rho$ be the quantum state $\Tensor_{(i,j) \in A} R_{ij}$, let $\Theta = (\Theta_i)_{i\in\Lambda}$ be a sequence of measurement bases chosen uniformly at random from $\prod_{i\in\Lambda} B_i$, and let $X = (X_i)_{i\in\Lambda}$ be the sequence of measurement outcomes when $\rho$ is measured in the bases $\Theta$ (i.e., each $X_i$ is the outcome of measuring subsystem $i$ in basis $\Theta_i$).  

Then, for any $\tau > 0$, and any $\lambda_i \in (0,\tfrac{1}{2})$ (for all $i\in\Lambda$), we have:
\begin{equation}
H_\infty^\eps(X|\Theta) \geq -\tau + \sum_{i\in\Lambda} (h_i - \lambda_i),
\end{equation}
where $\eps \leq \exp(-2\tau^2/c)$, and $c = \sum_{i\in\Lambda} 16\bigl(\lg\frac{|B_i|d_i}{\lambda_i}\bigr)^2$.

Now fix some small constants $0 < \lambda \ll \tfrac{1}{2}$ and $0 < \tau_0 \ll 1$.  Set $\lambda_i = \lambda$ (for all $i\in\Lambda$), and set $\tau = \tau_0 \sqrt{\ell/\lg q} \sqrt{c}$.  Then we have:
\begin{equation}
H_\infty^\eps(X|\Theta) \geq \tfrac{1}{2} \ell - \lambda \ell_0 - \tau,
\end{equation}
where $\eps \leq \exp(-2\tau_0^2 \ell/\lg q)$.  We can upper-bound $\tau$ as follows:
\begin{equation}
\begin{split}
\tau &= \tau_0 \sqrt{\ell/\lg q} \cdot 4 \Bigl( \sum_{i\in\Lambda} ( 1 + \ell_i + \lg\tfrac{1}{\lambda} )^2 \Bigr)^{1/2} \\
 &\leq \tau_0 \sqrt{\ell/\lg q} \cdot 4 \Bigl( (\sum_{i\in\Lambda} \ell_i^2)^{1/2} + (1+\lg\tfrac{1}{\lambda}) \sqrt{\ell_0} \Bigr) \\
 &\leq 4 \tau_0 \sqrt{\ell/\lg q} \, \Bigl( \sqrt{\lg q} \sqrt{\ell} + (1+\lg\tfrac{1}{\lambda}) \sqrt{\ell} \Bigr) \\
 &= 4 \tau_0 \ell \, \Bigl( 1 + \tfrac{1}{\sqrt{\lg q}} (1+\lg\tfrac{1}{\lambda}) \Bigr),
\end{split}
\end{equation}
where we used the triangle inequality for the $\ell_2$ norm, and the bounds $\ell_i \leq \lg q$, $\sum_{i\in\Lambda} \ell_i = \ell$ and $\ell_0 \leq \ell$.  Plugging this in above, and again using the bound $\ell_0 \leq \ell$, we get that:
\begin{equation}\label{eqn-puffin}
\begin{split}
H_\infty^\eps(X|\Theta)
 &\geq (\tfrac{1}{2}-\lambda) \ell - 4 \tau_0 \ell \, \Bigl( 1 + \tfrac{1}{\sqrt{\lg q}} (1+\lg\tfrac{1}{\lambda}) \Bigr).
\end{split}
\end{equation}

\subsubsection{Combining all the pieces}
\label{app-cherry}

First, we will show a lower-bound on $H_\infty^\eps(S,T|Q=0)$.  For any $s,t \in \set{0,1}^\ell$, we can upper-bound $P_{ST|Q}$ as follows:
\begin{equation}
\begin{split}
P_{ST|Q}(s,t|0)
&= \frac{P_{Q_B|STQ_A}(0|s,t,0) P_{ST|Q_A}(s,t|0)}{P_{Q_B|Q_A}(0|0)} \\
&\leq \frac{P_{ST|Q_A}(s,t|0)}{P_{Q_B|Q_A}(0|0)}
= P_{ST|Q_A}(s,t|0)\, \frac{\Pr[Q_A=0]}{\Pr[Q=0]}.
\end{split}
\end{equation}
From equation (\ref{eqn-treefungus}), we know that $\Pr[Q_A=0] = 2^{-\ell}$.  From the construction of the fictional adversary in section \ref{sec-dragonfly}, we know that $\Pr[Q=0] = \Pr[Z=z] / \Tr(M_z)$, where $Z$ is the output of the real adversary.  Finally, since $z$ was assumed to be ``non-negligible,'' we know that $\Pr[Z=z] / \Tr(M_z) \geq \delta / 2^{n\lg q}$.  Combining these facts, we get that
\begin{equation}
P_{ST|Q}(s,t|0) \leq P_{ST|Q_A}(s,t|0)\, \frac{2^{n\lg q}}{\delta\, 2^\ell},
\end{equation}
hence we conclude that
\begin{equation}\label{eqn-bee}
H_\infty^\eps(S,T|Q=0) \geq H_\infty^\eps(S,T|Q_A=0) - n\lg q + \ell - \lg(1/\delta).
\end{equation}

Note that the real adversary and the fictional adversary are related as follows (see equation (\ref{eqn-bluemold})):
\begin{equation}\label{eqn-ant}
H_\infty^\eps(S,T|Z=z) = H_\infty^\eps(S,T|Q=0).
\end{equation}
Combining equations (\ref{eqn-ant}), (\ref{eqn-bee}), (\ref{eqn-juniper3}) and (\ref{eqn-puffin}), we get that:
\begin{equation}
\begin{split}
H_\infty^\eps(S,T|Z=z)
&\geq H_\infty^\eps(S,T|Q_A=0) - n\lg q + \ell - \lg(1/\delta) \\
&\geq H_\infty^\eps(X|\Theta) + 2\ell - n\lg q - \lg\tfrac{1}{\delta} \\
&\geq (\tfrac{1}{2}-\lambda) \ell 
- 4\tau_0\ell\, \Bigl( 1 + \tfrac{1}{\sqrt{\lg q}} (1+\lg\tfrac{1}{\lambda}) \Bigr) 
+ 2\ell - n\lg q - \lg\tfrac{1}{\delta}.
\end{split}
\end{equation}

Note that the above bounds hold for any measurement outcome $z$ that is ``non-negligible.''  Moreover, the probability of observing a ``non-negligible'' measurement outcome is at least $1-\delta$ (by equation (\ref{eqn-greenmold})).  So we conclude that:
\begin{equation}
H_\infty^{\delta+\eps}(S,T|Z)
\geq (\tfrac{1}{2}-\lambda) \ell 
- 4\tau_0\ell\, \Bigl( 1 + \tfrac{1}{\sqrt{\lg q}} (1+\lg\tfrac{1}{\lambda}) \Bigr) 
+ 2\ell - n\lg q - \lg\tfrac{1}{\delta}.
\end{equation}

We can write this bound in a simpler form.  First, recall that we assumed the code $C$ has rate $\alpha>0$, i.e., $\ell \geq \alpha n\lg q$.  Then we have:
\begin{equation}
H_\infty^{\delta+\eps}(S,T|Z)
\geq \Bigl( (\tfrac{1}{2}-\lambda) 
- 4\tau_0\, ( 1 + \tfrac{1}{\sqrt{\lg q}} (1+\lg\tfrac{1}{\lambda}) ) 
+ (2-\tfrac{1}{\alpha}) \Bigr) \cdot \ell - \lg\tfrac{1}{\delta}.
\end{equation}
Typically we will let $\lambda$, $\tau_0$ and $\delta$ be small constants.  We will consider the asymptotic behavior as $\ell \rightarrow \infty$, we will let $q$ be large, and we will choose a family of codes $C$ that approaches the capacity of the $q$-ary symmetric channel; then $\alpha \approx \tfrac{1}{2}$.  Then we have the following bound:
\begin{equation}
H_\infty^{\delta+\eps}(S,T|Z)
\gtrsim \tfrac{1}{2} \ell.
\end{equation}


\section{Efficient implementations of one-time memories}

\subsection{Proof of Theorem \ref{thm-redtail}}
\label{app-saguaro}

First, we show that the code $C$ can transmit information reliably over the $q$-ary symmetric channel $\calE_q$ with error probability $p_e$.  Let us introduce some more random variables to describe the intermediate results of this process:
\begin{equation}
S \xrightarrow{\text{Encode } C_0} V \xrightarrow{\text{Encode } C_1} X 
\xrightarrow{\text{Channel } \calE_q} Z 
\xrightarrow{\text{Decode } C_1} B \xrightarrow{\text{Decode } C_0} \hat{S}.
\end{equation}
Here the message $S$ takes values in $\set{0,1}^{k\lg q_0}$, $V$ takes values in $\set{0,1}^{n\lg q_0}$, $X$ and $Z$ take values in $\set{0,1}^{n\lg q}$, $B$ takes values in $\bigl( \set{0,1}^{\lg q_0}\cup\set{*} \bigr)^n$, and $\hat{S}$ takes values in $\set{0,1}^{k\lg q_0}$.

Note that there are multiple sources of randomness in this picture:  the code $C$ is constructed using a random matrix $G_0$, the message $S$ is chosen at random, and the channel $\calE_q$ makes random errors.  We will use the following notation.  Expressions without subscripts, such as $\Pr[\hat{S} \neq S]$, denote probabilities summed over all possible choices of the message $S$ and all possible actions of the channel $\calE_q$; however, these expressions are still random variables that depend on the choice of the code $C$.  Expressions with a subscript $C$, such as $\Pr_C\bigl[ \Pr[\hat{S} \neq S] \geq \delta \bigr]$, denote probabilities summed over all possible choices of the code $C$.

First, consider the action of the channel $\calE_q$.  Let $N_e$ be the number of errors made by the channel (where each corrupted $q$-ary symbol counts as a single error), that is, 
\begin{equation}
N_e = |\set{i\in[n] \text{ s.t. } Z_i \neq X_i}|.  
\end{equation}
Note that $\EE N_e = np_e$.  Choose any constant $\tau$ such that $0<\tau<\theta$.  Define $r := n(p_e+\tau)$, and note that by Hoeffding's inequality, $\Pr[N_e > r] \leq e^{-2\tau^2 n}$.  

Now consider the decoding algorithm for the code $C_1$.  Let $N_{ude}$ be the number of errors that are not detected by $C_1$, that is, 
\begin{equation}
N_{ude} = |\set{i\in[n] \text{ s.t. } B_i \neq * \text{ and } B_i \neq V_i}|.  
\end{equation}
Note that, for any $i\in[n]$, we have $\Pr[B_i \neq * \text{ and } B_i \neq V_i] = p_e(q_0-1)/(q-1)$.  Using the union bound, we have that $\Pr[N_{ude} > 0] \leq np_e(q_0-1)/(q-1)$.  Finally, using equations (\ref{eqn-def-q}) and (\ref{eqn-def-q0}), note that 
\begin{equation}
\frac{q_0-1}{q-1} \leq \frac{q_0}{q} = \frac{1}{2^{c-c_0}}, \quad
c-c_0 \geq \eps n + \lg(np_e).
\end{equation}
Combining these facts, we get the bound $\Pr[N_{ude} > 0] \leq 2^{-\eps n}$.

Thus we can write:
\begin{equation}\label{eqn-coral}
\begin{split}
\Pr[\hat{S} \neq S] 
&\leq \Pr[N_e > r \text{ or } N_{ude} > 0] 
+ \Pr[\hat{S} \neq S \text{ and } N_e \leq r \text{ and } N_{ude} = 0] \\
&\leq e^{-2\tau^2 n} + 2^{-\eps n} 
+ \Pr[\hat{S} \neq S \text{ and } N_e \leq r \text{ and } N_{ude} = 0].
\end{split}
\end{equation}

Now consider the case where $N_e \leq r$ and $N_{ude} = 0$.  We will analyze the decoding process for the code $C_0$.  Look at the random variable $B = (B_1,B_2,\ldots,B_n)$, which is the input to the decoder.  We know that at most $r$ of the coordinates $B_i$ are $*$ symbols, and those coordinates $B_i$ that are not $*$ symbols must be equal to the corresponding coordinates $V_i$.  We introduce some notation:  for any $b \in \bigl( \set{0,1}^{\lg q_0} \cup \set{*} \bigr)^n$, let us define $C_0^{-1}(b)$ to be the set of all possible messages that are consistent with $b$, that is, 
\begin{equation}
C_0^{-1}(b) = \set{t \in \set{0,1}^{k\lg q_0} \text{ such that, } \forall i\in[n] \text{ with } b_i\neq*,\, 
\forall j\in[\lg q_0],\, C_0(t)_{ij} = b_{ij}}.
\end{equation}
The decoding algorithm for $C_0$ will search for any message in the set $C_0^{-1}(B)$.  Note that the correct message $S$ lies inside $C_0^{-1}(V)$, which is contained in $C_0^{-1}(B)$.  A decoding failure $\hat{S} \neq S$ implies that there must exist some other message $t \in C_0^{-1}(B)$ such that $t \neq S$.  

So we can write:
\begin{equation}\label{eqn-kelp}
\begin{split}
\Pr[&\hat{S} \neq S \text{ and } N_e \leq r \text{ and } N_{ude} = 0] \\
&\leq \Pr[(\exists t \in C_0^{-1}(B) \text{ s.t. } t \neq S) \text{ and } N_e \leq r \text{ and } N_{ude} = 0] \\
&\leq \Pr[\exists t \in C_0^{-1}(B) \text{ s.t. } t \neq S \;|\; N_e \leq r] \\
&= 2^{-k\lg q_0} \sum_{s\in\set{0,1}^{k\lg q_0}} \Pr[\exists t \in C_0^{-1}(B) \text{ s.t. } t \neq S \;|\; S=s \text{ and } N_e \leq r] \\
&\leq 2^{-k\lg q_0} \sum_{s\in\set{0,1}^{k\lg q_0}} \sum_{t\in\set{0,1}^{k\lg q_0}\setminus\set{s}} \Pr[t \in C_0^{-1}(B) \;|\; S=s \text{ and } N_e \leq r] \\
&= 2^{-k\lg q_0} \sum_{s\in\set{0,1}^{k\lg q_0}} \sum_{t\in\set{0,1}^{k\lg q_0} \setminus \set{s}} \sum_{\substack{b\in(\set{0,1}^{\lg q_0}\cup\set{*})^n\\ |\set{i\in[n] \text{ s.t. } b_i=*}| \leq r}} 1[t \in C_0^{-1}(b)] \Pr[B=b \;|\; S=s \text{ and } N_e \leq r], 
\end{split}
\end{equation}
where in the third step we used the fact that the number of errors $N_e$ made by the channel $\calE_q$ is independent of the message $S$.

We now calculate the expectation value of this quantity, averaging over the random choice of the code $C$ (that is, the random choice of the matrix $G_0$).  Note that $1[t \in C_0^{-1}(b)]$ is a random variable that depends on $C_0(t) = t^T G_0$, and $\Pr[B=b \;|\; S=s \text{ and } N_e \leq r]$ is a random variable that depends on $C_0(s) = s^T G_0$.  Note that $s \neq t$ implies that $s$ and $t$ are linearly independent (since $s$ and $t$ are vectors over $GF(2)$); hence $s^T G_0$ and $t^T G_0$ are independent random variables.  So we can write:
\begin{equation}
\EE_C\bigl[ 1[t \in C_0^{-1}(b)] \Pr[B=b \;|\; S=s \text{ and } N_e \leq r] \bigr] 
= \EE_C\bigl[ 1[t \in C_0^{-1}(b)] \bigr] \EE_C\bigl[ \Pr[B=b \;|\; S=s \text{ and } N_e \leq r] \bigr].
\end{equation}
We can bound the first of these two factors as follows:
\begin{equation}
\begin{split}
\EE_C\bigl[ &1[t \in C_0^{-1}(b)] \bigr] = \Pr_C[t \in C_0^{-1}(b)] \\
&= \Pr_C[\forall i\in[n] \text{ with } b_i\neq*,\, \forall j\in[\lg q_0],\, C_0(t)_{ij} = b_{ij}] \\
&= 2^{-|\set{i\in[n] \text{ s.t. } b_i\neq*}| \cdot \lg q_0} \leq 2^{-(n-r) \lg q_0}.
\end{split}
\end{equation}

Substituting into equation (\ref{eqn-kelp}), and using the bound $k \leq n(1-p_e-\theta) + 1$ from (\ref{eqn-def-n}), we get that:
\begin{equation}
\begin{split}
\EE_C \Pr[&\hat{S} \neq S \text{ and } N_e \leq r \text{ and } N_{ude} = 0] \\
&\leq 2^{-k\lg q_0} \sum_{s\in\set{0,1}^{k\lg q_0}} \sum_{t\in\set{0,1}^{k\lg q_0} \setminus \set{s}} \sum_{\substack{b\in(\set{0,1}^{\lg q_0}\cup\set{*})^n\\ |\set{i\in[n] \text{ s.t. } b_i=*}| \leq r}} 2^{-(n-r) \lg q_0} \EE_C\bigl[ \Pr[B=b \;|\; S=s \text{ and } N_e \leq r] \bigr] \\
&< 2^{k\lg q_0} 2^{-(n-r) \lg q_0} \\
&\leq 2^{(n(1-p_e-\theta) + 1 - n + n(p_e+\tau)) \lg q_0} \\
&= 2^{(-n\theta+n\tau+1)\lg q_0}.
\end{split}
\end{equation}
Plugging into equation (\ref{eqn-coral}), we get:
\begin{equation}
\EE_C \Pr[\hat{S} \neq S] \leq e^{-2\tau^2 n} + 2^{-\eps n} + 2^{(-n\theta+n\tau+1)\lg q_0}.
\end{equation}
Finally, Markov's inequality implies that, for any $\lambda \gg 1$, 
\begin{equation}
\Pr_C\Bigl[ \Pr[\hat{S} \neq S] \geq \lambda \bigl( e^{-2\tau^2 n} + 2^{-\eps n} + 2^{(-n\theta+n\tau+1)\lg q_0} \bigr) \Bigr]
\leq \frac{1}{\lambda}.
\end{equation}
This proves the first part of Theorem \ref{thm-redtail}.

We now show the remaining parts of the theorem.  It is clear from the construction that $C$ is a linear code over $GF(2)$.  The rate of the code $C$ can be bounded as follows, using equations (\ref{eqn-def-q}) and (\ref{eqn-def-q0}):
\begin{equation}
\alpha := \frac{k\lg q_0}{n\lg q} \geq (1-p_e-\theta) \frac{\lg q_0}{\lg q} 
= (1-p_e-\theta) \frac{c_0}{c} \geq (1-p_e-\theta) (1-\delta).
\end{equation}
Finally, note that the encoding procedure for $C$ consists of matrix multiplications over $GF(2)$, while the decoding procedure can be implemented by solving linear systems of equations over $GF(2)$; hence both procedures take time polynomial in $n\lg q$.  (Also note that $\lg q$ grows at most linearly with $n$, and $n$ is proportional to $k$.)  This completes the proof of Theorem \ref{thm-redtail}.


\end{document}